\documentclass[aps,pra,showpacs,amssymb,twocolumn]{revtex4}
\usepackage{graphicx}
\usepackage{epstopdf} 
\DeclareGraphicsExtensions{.pdf,.eps,.png,.jpg,.mps}
\usepackage{hyperref}
\usepackage{bm,amsmath}
\usepackage{dcolumn}

\newcommand{\bra}[1]{\left\langle #1 \right|}
\newcommand{\ket}[1]{\left|#1\right\rangle}

\begin{document}
\title{Controlled-\small{NOT} gate with weakly coupled qubits: Dependence of fidelity on the form of interaction}

\author{Joydip Ghosh}
\email{jghosh@physast.uga.edu}
\author{Michael R. Geller}
\email{mgeller@uga.edu}
\affiliation{Department of Physics and Astronomy, University of Georgia, Athens, GA 30602, USA}
\date{\today}

\begin{abstract}

An approach to the construction of the CNOT quantum logic gate for a 4-dimensional coupled-qubit model with weak but otherwise arbitrary coupling has been given recently [M.~R. Geller \textit{et al.}, Phys. Rev. A {\bf 81}, 012320 (2010)]. How does the resulting fidelity depend on the form of qubit-qubit coupling? In this paper we calculate intrinsic fidelity curves (fidelity in the absence of decoherence versus total gate time) for a variety of qubit-qubit interactions, including the commonly occurring isotropic Heisenberg and XY models, as well as randomly generated ones. For interactions not too close to that of the Ising model, we find that the fidelity curves do not significantly depend on the form of the interaction, and we calculate the resulting interaction-averaged fidelity curve for the non-Ising-like cases and a criterion for determining its applicability.

\end{abstract}

\maketitle

\section{Introduction}

The operation of a quantum computer requires the implementation of a universal two-qubit quantum logic gate, such as the CNOT \cite{ncbook}. The problem of constructing a CNOT gate has been addressed from various perspectives and for different physical systems \cite{PhysRevLett.74.4091,lloyd2002qip,ZhangPRA03,GaliautdinovPRA07,GaliautdinovJMP07,GaliautdinovPRA09,Monroe1995prl,blatt2003nature,obrien2003nature,PhysRevLett.93.020504,Mooij2007nature}. In recent work, Geller \textit{et al.} \cite{PhysRevA.81.012320} approached the problem of CNOT construction from a somewhat general standpoint. Starting with a Hamiltonian for a four-dimensional coupled-qubit model, they derived a CNOT pulse sequence assuming weak coupling. In this work, we calculate the intrinsic fidelity of the CNOTs constructed according to this protocol.

The question of CNOT gate fidelity has already been discussed from other standpoints \cite{hill2007prl,felloni2009arxiv}. Fidelity loss can be separated into intrinsic errors, which include errors resulting from the use of weak-coupling and weak-driving approximations, and errors resulting from decoherence \cite{RevModPhys.75.715,motzoi2009prl}. The latter, which of course depends sensitively on the experimental architecture and noise sources, is largely a function of total gate time. Therefore, by evaluating the intrinsic fidelity as a function of total gate time (which indirectly determines the strength of the qubit-qubit interaction), we can separate intrinsic and extrinsic errors in a way that allows application to a wide variety of architectures and environments. The CNOT fidelity curves we calculate are the intrinsic fidelities as a function of gate time, optimized over all pulse sequence parameters and coupling constants that lead to the same total gate time. We calculate fidelity curves for coupled-qubit models with commonly occurring forms of interaction, such as XY and Heisenberg couplings, as well as for randomly generated ones with lower symmetry. We find that for
qubit-qubit interactions  not too close to that of the Ising model, the fidelity curves are largely insensitive to 
the form of the interaction. This allows us to provide a single fidelity curve for non-Ising-like models and a
criterion for determining its applicability.

The remainder of the paper is organized as follows: In Sec.~\hyperref[sec2]{\ref{sec2}}, we review the perturbative CNOT gate design derived in Ref.~\cite{PhysRevA.81.012320} and describe the model considered there. 
In Sec.~\hyperref[sec3]{\ref{sec3}}, we discuss the various sources of intrinsic errors and define the fidelity
measure used for all subsequent computations. In Sec.~\hyperref[sec4]{\ref{sec4}}, we explain the methodology used for our fidelity calculations, which involve exact numerical simulations of the underlying coupled-qubit models. In Sec.~\hyperref[sec5]{\ref{sec5}}, we present fidelity curves for different forms of interaction and the interaction-averaged fidelity, and in Sec.~\hyperref[sec6]{\ref{sec6}} we explain the 
poor performance when the interaction is close to that of the Ising model. 

\section{CNOT Protocol} \label{sec2}
In this section, we briefly review the main results of Ref.~\cite{PhysRevA.81.012320} for constructing a CNOT
gate pulse sequence.

\subsection{Model Hamiltonian} \label{sec2.1}

The Hamiltonian for a wide variety of physical systems being considered for quantum computation can be written as
\begin{equation} \label{hamiltonian}
\begin{split}
H=\sum_{i=1,2}\left[-\frac{\epsilon_i}{2}{\sigma}^{z}_{i}+\Omega_{i}\cos\left(\frac{\epsilon_{i}t}{\hbar}+\phi_i\right){\sigma}^{x}_{i}\right] \\ +\sum_{\mu,\nu=x,y,z}J_{\mu\nu}\;\sigma^{\mu}_{1}\otimes\sigma^{\nu}_{2},
\end{split}
\end{equation}
where $J_{\mu\nu}$ is a $3\times3$ coupling matrix which takes different forms for different architectures under consideration. The parameters $\epsilon_i$ and $\Omega_i$ (with $\Omega_i \ll \epsilon_i$) are  tunable and weak coupling ($|J_{\mu\nu}| \ll \epsilon_i$) is assumed. The CNOT gates are implemented according to a pulse sequence consisting of two entangling operations along with single qubit rotations. The entangling operations are carried out with tuned qubits ($\epsilon_1=\epsilon_2$) and the local rotations are performed with detuned qubits. Table~\ref{paramtable} gives values of the model parameters used for our simulations. For weakly coupled tuned qubits, the Hamiltonian{\eqref{hamiltonian}} can be written in the interaction picture (or rotating frame) as

\begin{equation} \label{hamiltonianrwa}
\begin{split}
H \approx \sum_{i=1,2}\frac{\Omega_i}{2}\left(\cos\phi_i \, \sigma_{i}^{x}-\sin\phi_i \, 
\sigma_{i}^{y}\right)+{\cal H},
\end{split}
\end{equation}
where
\begin{equation} \label{hamiltonianint}
{\cal H}\equiv J\left(\sigma_{1}^{x}\sigma_{2}^{x}+\sigma_{1}^{y}\sigma_{2}^{y}\right)+J_{zz}\sigma_{1}^{z}\sigma_{2}^{z}+J'\left(\sigma_{1}^{x}\sigma_{2}^{y}-\sigma_{1}^{y}\sigma_{2}^{x}\right).
\end{equation}
The parameters $J$ and $J'$ in Eq.{\eqref{hamiltonianint}} are given by
\begin{equation} \label{Jdef}
J\equiv\frac{J_{xx}+J_{yy}}{2} \ \ \mbox{and}\ \  J'\equiv\frac{J_{xy}-J_{yx}}{2}.
\end{equation}

\begin{table}[htb]
\centering
\caption{\label{paramtable}Parameter values used in this work. The tuned qubit frequency is the frequency
of the qubits used during the entangling operations. During single-qubit operations, the frequency of qubit 2 is suddenly increased to 11 GHz. The ranges of allowed values of Rabi frequencies and overall coupling strengths are used to constrain the optimization described in the body of the paper. $J_{\mu\nu}^{*}$ and $g$ are defined in Eq.~\eqref{jbasisdef}.}
\begin{tabular}{||c | c||}
\hline & \\
parameter & value\\
\hline & \\
common tuned qubit frequency $\epsilon/h$  & 10 GHz \\
qubit-qubit detuning  & 1 GHz \\
range of allowed Rabi frequencies $\Omega/h$ & 50-500 \rm{MHz} \\
range of allowed coupling strengths $g/h$ & 1-500 \rm{MHz} \\
range of gate times $t_{\rm gate}$ considered  & 10-50 ns \\
& \\
$J_{\mu\nu}^{*}$ for Isotropic Heisenberg coupling & $\begin{bmatrix} 1&0&0\\ 0&1&0\\ 0&0&1 \end{bmatrix}$ \\
& \\
$J_{\mu\nu}^{*}$ for Ising coupling & $\begin{bmatrix} 0&0&0\\ 0&0&0\\ 0&0&1 \end{bmatrix}$ \\
& \\
$J_{\mu\nu}^{*}$ for XY coupling & $\begin{bmatrix} 1&0&0\\ 0&1&0\\ 0&0&0 \end{bmatrix}$ \\
& \\
\hline
\end{tabular}
\end{table}

\subsection{CNOT pulse sequence} \label{sec2.2}
The pulse sequence derived in Ref.\cite{PhysRevA.81.012320}, carried out from right to left, is
\begin{equation} \label{gellercnotps}
\begin{split}
\mbox{CNOT}=e^{i\frac{3\pi}{4}}R_{y}\left(-\frac{\pi}{2}\right)_{1}R_{x}\left(-\frac{\pi}{2}\right)_{2}R_{z}\left(-\varphi\right)_{2}\\ \times R_{x}\left(\frac{\pi}{2}\right)_{1}e^{-i{\cal H}\Delta t/\hbar}R_{x}\left(\pi\right)_{1}e^{-i{\cal H}\Delta t/\hbar}R_{z}\left(\varphi\right)_{2}R_{y}\left(\frac{\pi}{2}\right)_{1},
\end{split}
\end{equation}
where $R_{\mu}(\theta)_{i} \equiv e^{-\frac{i}{2}(\theta)\sigma_{i}^{\mu}}$ (with $\mu=x,y,z$ and $i$=1,2) is a single qubit rotation. Here
\begin{equation} \label{phidef}
\varphi \equiv \arg\left(J+iJ'\right) \  \ \mbox{and} \ \  \Delta t \equiv \frac{\pi \hbar}{8\sqrt{J^{2}+J'^{2}}}.
\end{equation}
The operator $e^{-i{\cal H}\Delta t/\hbar}$ represents the action of bringing the qubits into resonance for a time $\Delta t$. The CNOT pulse sequence given in Eq.{\eqref{gellercnotps}} involves two rotations about the $z$ axis. For our exact simulations below, it will be convenient to rewrite (\ref{gellercnotps}) in terms of $x$ and $y$ rotations, leading to
\begin{equation} \label{finalps}
\begin{split}
\mbox{CNOT}=e^{i\frac{3\pi}{4}}\left[R_{y}\left(\frac{\pi}{2}\right)_{1}R_{x}\left(-\frac{\pi}{2}\right)_{2}\right]R_{y}\left(\varphi\right)_{2}R_{x}\left(\frac{\pi}{2}\right)_{2}\\
\quad \times e^{-i{\cal H}\Delta t/\hbar}R_{x}\left(\pi\right)_{1}e^{-i{\cal H}\Delta t/\hbar}\left[R_{x}\left(\frac{\pi}{2}\right)_{1}R_{x}\left(-\frac{\pi}{2}\right)_{2}\right]\\ \times R_{y}\left(-\varphi\right)_{2}R_{y}\left(\frac{\pi}{2}\right)_{1}.
\end{split}
\end{equation}
This is the CNOT pulse sequence used in the present analysis. Operations inside square brackets can be performed simultaneously.

\section{Intrinsic Fidelity} \label{sec3}

The CNOT pulse sequence (\ref{finalps}) is an exact identity; the errors come from realizing the 
individual terms in it using Hamiltonian (\ref{hamiltonian}). Here we discuss the possible sources 
of errors and the precise definition 
of fidelity used in this work.

\subsection{Sources of error} \label{sec3.1}

As mentioned earlier, we are not concerned here with extrinsic errors, such as noise and decoherence, 
since these depend very much on the specific experimental architecture and noise sources. The fidelity loss computed here originates from intrinsic sources. The exact CNOT pulse sequence \eqref{finalps} is derived in the context of the approximate Hamiltonian \eqref{hamiltonianrwa}, which is derived from the model Hamiltonian \eqref{hamiltonian} assuming weak coupling ($|J_{\mu\nu}| \ll \epsilon_i$) and weak driving
 ($\Omega_i \ll \epsilon_i$). These approximations contribute to the accumulation of fidelity loss. In addition, we also assume that the coupling is always on, even when the local rotations are being performed. This assumption is necessitated by the fixed coupling used in most experimental architectures. Due to the presence of more local rotations than entangling operations in the pulse sequence, the latter causes a larger contribution to the intrinsic error.

\subsection{Definition of fidelity} \label{sec3.2}

In general, fidelity gives a measure of how close two quantum states are. There exists different measures of fidelity. The definition we adopt is given by (See Ref.\cite{preskillbook} Page 222, Eq.14)
\begin{equation} \label{fdef}
F\left(\ket{\psi},\rho\right) \equiv \bra{\psi}\rho\ket{\psi},
\end{equation}
where $\ket{\psi}$ is considered to be a pure state and $\rho$ is the density matrix of an arbitrary state. Here we are interested in calculating fidelity between two operations (ideal CNOT and realized CNOT) which requires some modifications of the definition given by Eq.{\eqref{fdef}}. In this context fidelity means how close these operations are. Suppose, we have two unitary operations $U$ and $U_{\rm target}$ and we want to calculate the fidelity between these operations. A natural way to understand this closeness is to take a randomly generated vector $\ket{\chi}$ (defined on a Hilbert space), then apply the operations $U$ and $U_{\rm target}$ on the vector to obtain transformed vectors $U\ket{\chi}$ and $U_{\rm target}\ket{\chi},$ and finally identify these transformed states with $\rho$ and $\ket{\psi}$ in Eq.{\eqref{fdef}} to derive an 
expression for fidelity that depends on the state $\ket{\chi}$,
\begin{equation} \label{fchidef}
F_{\chi}\left(U_{\rm target},U\right)={\underbrace{\bra{\chi}U_{\rm target}^{\dagger}}_{\bra{\psi}}\underbrace{U\ket{\chi}\bra{\chi}U^{\dagger}}_{\rho}\underbrace{U_{\rm target}\ket{\chi}}_{\ket{\psi}}}.
\end{equation}
Finally, we average over randomly generated $\ket{\chi}$ (chosen from a uniform distribution) to introduce an average fidelity, according to
\begin{equation} \label{favgdefold}
F_{\rm average}\left(U_{\rm target},U\right)=\frac{1}{N\left(\ket{\chi}\right)}\sum_{\ket{\chi}}F_{\chi}\left(U_{\rm target},U\right),
\end{equation}
where $N\left(\ket{\chi}\right)$ is the total number of randomly generated $\ket{\chi}$ states. To obtain a closed form expression of fidelity we change this sum to an integral, 
\begin{equation} \label{favgintdef}
F_{\rm average}\left(U_{\rm target},U\right)=\int \left|\bra{\chi}{\cal M}\ket{\chi}\right|^{2} dV,
\end{equation}
where ${\cal M} \equiv U_{\rm target}^{\dagger}U$ and $dV$ is a normalized measure. It has already been proven \cite{PhysRevA.70.012315,Pedersen200747,Pedersen20087028} that, for any linear operator $M$ on an $n$-dimensional complex Hilbert space,
\begin{equation} \label{ftracedef}
\displaystyle\int\limits_{S^{2n-1}} \left|\bra{\chi}M\ket{\chi}\right|^{2} dV = \frac{{\rm Tr}(MM^{\dagger})+\left|{\rm Tr}(M)\right|^{2}}{n(n+1)},
\end{equation}
where the normalized state vectors $\ket{\chi}$ are defined on the unit sphere $S^{2n-1}$ in $\mathbb{C}$. Using Eq.{\eqref{ftracedef}} for a 4-dimensional Hilbert space, we can rewrite our expression for average fidelity as
\begin{equation} \label{favgdef}
F_{\rm average}\left(U_{\rm target},U\right) = \frac{4+\left|{\rm Tr} \, (U_{\rm target}^{\dagger}U)\right|^{2}}{20}.
\end{equation}
We use {\eqref{favgdef}} for computing fidelity between any two unitary quantum operations and express it in percent.

\section{Simulations} \label{sec4}

For a given qubit-qubit coupling tensor $J_{\mu\nu}$, the pulse sequence (\ref{finalps}) is realized by performing the pair of two-qubit entangling operations with tuned qubits [for a time $\Delta t$ given in (\ref{phidef})] and the single-qubit operations with strongly detuned qubits. The time to implement the full pulse sequence depends on  $J_{\mu\nu}$ and the Rabi frequencies, which in principle can be different for each qubit and for each local rotation required. However, in this work we choose all Rabi frequencies to have the same value. 

The coupling tensor $J_{\mu\nu}$ can be decomposed according to
\begin{equation} \label{jbasisdef}
J_{\mu\nu} = g\times J_{\mu\nu}^{*},
\end{equation}
where $g>0$ is a measure of the overall strength and $J_{\mu\nu}^{*}$ describes the form
of the coupling. $J_{\mu\nu}^{*}$ is defined to satisfy
\begin{equation}
|J_{\mu\nu}^{*}| \leq 1 \ \ \ \ \ {\rm all} \ \mu, \nu.
\end{equation}
Three important examples of  $J_{\mu\nu}^{*}$ are given in Table~\ref{paramtable}.

The simulations reported here are obtained by exact numerical integration of the model (\ref{hamiltonian}).Our choice of fixed experimental parameters was motivated by superconducting architectures \cite{Geller2007Springer1}. We optimize over Rabi frequencies and qubit-qubit interaction strengths only, and do not allow for variation in the local rotation angles of Eq.(\ref{finalps}). Although small refinements of these rotation angles can make slight improvements in the fidelity (by compensating for the qubit coupling that is suppressed by detuning but still always present), the fidelity changes are small on the scale of the main effects we consider (the dependence on the form of qubit-qubit interaction). These considerations lead us to vary the coupling tensor $J_{\mu\nu}$, total gate time and Rabi frequency and compute fidelity as a function of these variables. But since we know the pulse sequence, we can determine the total gate time  as a function of $J_{\mu\nu}$ and the single Rabi frequency by adding up the time required for each operation. So, for the simulation we fix total gate time, vary Rabi frequency within an allowed range given in Table~\ref{paramtable}, compute corresponding values of $g$ and optimize the fidelity from the evolution of the original Hamiltonian(\ref{hamiltonian}). This procedure leads to a single point on a fidelity curve.

\section{Fidelity curves} \label{sec5}

\begin{figure}[htb]
	\centering
			\includegraphics[width=8.0cm]{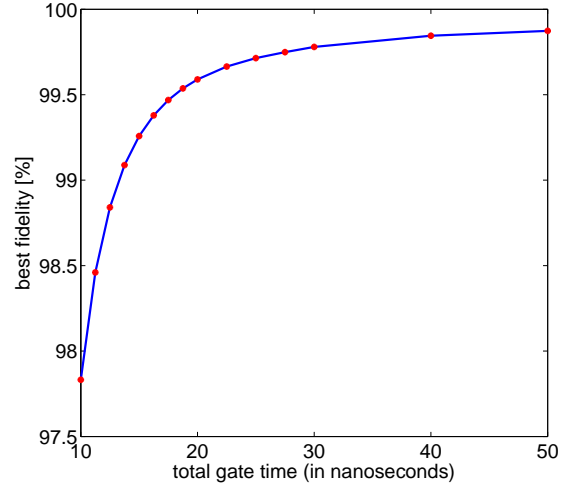}
				\caption{(Color online) Plot of optimal  fidelity versus total CNOT gate time for the Heisenberg interaction.}
		\label{heisen}
\end{figure}

\begin{figure}[htb]
	\centering
			\includegraphics[width=8.0cm]{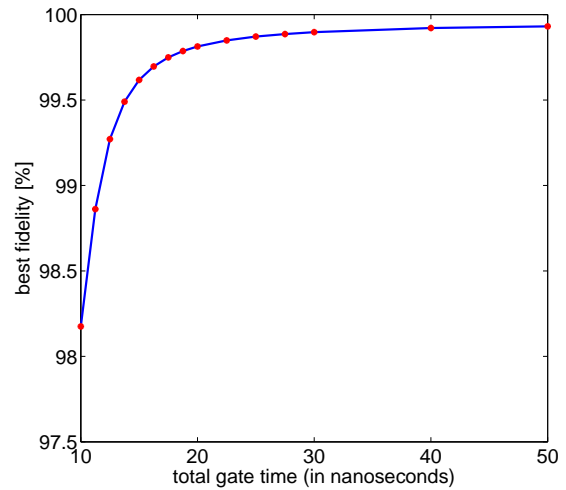}
				\caption{(Color online) Optimal fidelity versus total gate time for the XY interaction.}
		\label{XY}
\end{figure}

Figures \ref{heisen} and \ref{XY} give the optimal CNOT fidelities as a function of total gate time for
the Heisenberg and XY interactions and Tables \ref{heisentable} and \ref{xytable} show corresponding optimal values of relevant parameters. The fidelity curves are similar, indicating that a fidelity of
99\% can be obtained in less than 15 ns and 99.9\% can be obtained in about 50 ns.  Alternatively,
these results indicate that for these common forms of qubit-qubit coupling, 99\% can be achieved
with a coherence time in excess of 15 ns and 99.9\% can be achieved with at least 50 ns of
coherence. [We remind the reader that the model (\ref{hamiltonian}) does not include higher energy (non-qubit) states, which further limit performance, and that results are obtained for 10 GHz qubits.]

\begin{table}[htb]
\centering
\caption{\label{heisentable}Optimum values of parameters for Heisenberg interaction.}
\begin{tabular}{||c | c | c | c ||}
\hline & \ & \ & \\
total time (ns) & fidelity[$\%$] & $g/h$ (\rm{MHz}) & $\Omega/h$ (\rm{MHz})\\
\hline & \ & \ & \\
	 10.00 &  97.8321 &  19.1964 & 430 \\
   11.25 &  98.4599 &  16.1049 & 430 \\
   12.50 &  98.8405 &  13.8710 & 430 \\
   13.75 &  99.0881 &  12.1813 & 430 \\
   15.00 &  99.2579 &  10.8586 & 430 \\
   16.25 &  99.3792 &   9.7950 & 430 \\
   17.50 &  99.4688 &   8.9212 & 430 \\
   18.75 &  99.5368 &   8.1905 & 430 \\
   20.00 &  99.5895 &   7.5704 & 430 \\
   22.50 &  99.6646 &   6.5749 & 430 \\
   25.00 &  99.7144 &   5.8108 & 430 \\
   27.50 &  99.7489 &   5.2058 & 430 \\
   30.00 &  99.7794 &   4.8851 & 340 \\
   40.00 &  99.8452 &   3.5124 & 340 \\
   50.00 &  99.8734 &   2.7419 & 340 \\
& \ & \ & \\
\hline
\end{tabular}
\end{table}

\begin{table}[htb]
\centering
\caption{\label{xytable}Optimum values of parameters for XY interaction.}
\begin{tabular}{||c | c | c | c ||}
\hline & \ & \ & \\
total time (ns) & fidelity[$\%$] & $g/h$ (\rm{MHz}) & $\Omega/h$ (\rm{MHz})\\
\hline & \ & \ & \\
	 10.00 &  98.1750 &  17.8571 & 500 \\
   11.25 &  98.8618 &  23.8095 & 250 \\
   12.50 &  99.2710 &  19.2308 & 250 \\
   13.75 &  99.4902 &  16.6667 & 240 \\
   15.00 &  99.6174 &  14.2857 & 240 \\
   16.25 &  99.6966 &  12.5000 & 240 \\
   17.50 &  99.7494 &  11.1111 & 240 \\
   18.75 &  99.7864 &  10.0000 & 240 \\
   20.00 &  99.8133 &   9.0909 & 240 \\
   22.50 &  99.8491 &   7.6923 & 240 \\
   25.00 &  99.8713 &   6.6667 & 240 \\
   27.50 &  99.8861 &   5.8824 & 240 \\
   30.00 &  99.8973 &   5.2083 & 250 \\
   40.00 &  99.9211 &   3.6765 & 250 \\
   50.00 &  99.9311 &   2.8409 & 250 \\
& \ & \ & \\
\hline
\end{tabular}
\end{table}

The weak dependence of the fidelity curve on the form of interaction is typical, unless the interaction is
close to that of the Ising model (see Table~\ref{paramtable}). To quantify
this closeness we define a parameter [recall (\ref{Jdef}) and (\ref{jbasisdef})]
\begin{equation} \label{etadef}
\eta \equiv \frac{\sqrt{J^{2}+J'^{2}}}{g}.
\end{equation}
It can be shown that $0 \leq \eta \leq \sqrt{2}$. For the Ising interaction, $\eta = 0$, whereas for the
Heisenberg and XY interactions, $\eta = 1$. A ``typical'' value of $\eta$, defined by averaging
the function $\eta(J_{\mu \nu})$ over an unconstrained uniform distribution of $J_{\mu \nu}$ tensors, is about 
$0.52.$

\begin{figure}[htb]
	\centering
		\includegraphics[width=8.0cm]{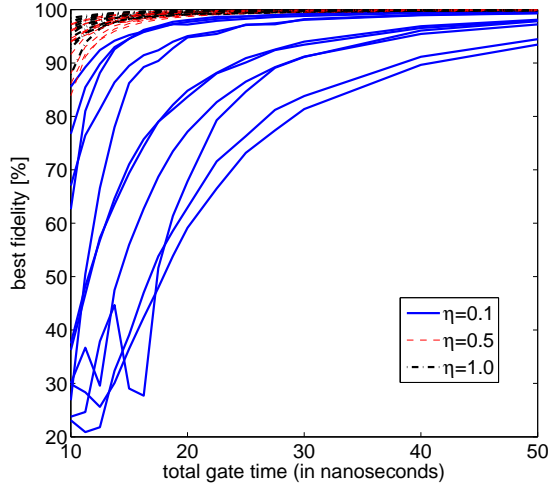}
		\caption{(Color online) Plot of optimal fidelity versus total gate time for random interactions.}
	\label{random}
\end{figure}

In Fig.~\ref{random} we show fidelity curves for randomly generated forms of interaction with three 
fixed values $\eta$. The unambiguous loss of fidelity for $\eta=0.1$ is due to the fact that the interaction 
is close to the Ising limit ($\eta=0$).  The similar behavior of fidelity for $\eta=0.5$ and $\eta=1.0$ affirms our assertion that fidelity curves do not significantly depend on the form of the interaction as long
as $\eta$ is not too close to zero. The reason for the poor performance for small $\eta$ is discussed in 
Sec.~\ref{sec6}.

\begin{figure}[htb]
	\centering
		\includegraphics[width=8.0cm]{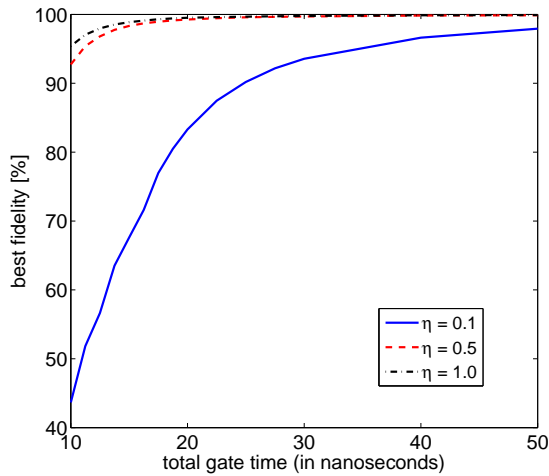}
		\caption{(Color online) Plot of optimal fidelity versus total gate time, averaged over randomly
		generated interactions with fixed values of $\eta$, for $\eta=0.1, 0.5,$ and $1.0$.}
	\label{randomavg}
\end{figure}

Given that the fidelity is largely independent of the form of interaction, as long as $\eta$ is not too small,
it is useful to average over interaction forms to obtain interaction-independent fidelity curves. This is
provided in Fig.~\ref{randomavg}, which present interaction-averaged fidelity curves for $\eta=0.1, 0.5,$ and $1.0$.

\section{Conclusions} \label{sec6}

We have shown that the intrinsic fidelity versus total gate time for the CNOT gate as constructed 
by Eq.~(\ref{finalps}) is largely independent of the form of qubit-qubit interaction, as long as that
interaction is not too close to that of the Ising model, as measured by the the parameter $\eta$
defined in (\ref{etadef}). For typical (non-Ising-like) couplings, the fidelity is given in
Fig.~\ref{randomavg}); here one can use either the $\eta=0.5$ or $\eta=1.0$ curves.

\begin{table}[h]
\centering
\caption{\label{heisendecohtable}Optimum values of parameters for Heisenberg interaction with 500 ns. amplitude damping.}
\begin{tabular}{||c | c | c | c ||}
\hline & \ & \ & \\
total time (ns) & fidelity[$\%$] & $g/h$ (\rm{MHz}) & $\Omega/h$ (\rm{MHz})\\
\hline & \ & \ & \\
	 10.00 &  96.9272 &  21.0227 & 370 \\
   11.25 &  97.3098 &  17.3709 & 370 \\
   12.50 &  97.4493 &  14.8000 & 370 \\
   13.75 &  97.4699 &  12.8920 & 370 \\
   15.00 &  97.4260 &  11.4198 & 370 \\
   16.25 &  97.3423 &  10.2493 & 370 \\
   17.50 &  97.2355 &   9.2965 & 370 \\
   18.75 &  97.1125 &   8.5057 & 370 \\
   20.00 &  96.9781 &   7.8390 & 370 \\
   22.50 &  96.6904 &   6.7766 & 370 \\
   25.00 &  96.3866 &   5.9677 & 370 \\
   27.50 &  96.0740 &   5.3314 & 370 \\
   30.00 &  95.7565 &   4.8177 & 370 \\
   40.00 &  94.4736 &   3.4774 & 370 \\
   50.00 &  93.1969 &   2.7206 & 370 \\
& \ & \ & \\
\hline
\end{tabular}
\end{table}

\begin{figure}[h]
	\centering
		\includegraphics[width=8.0cm]{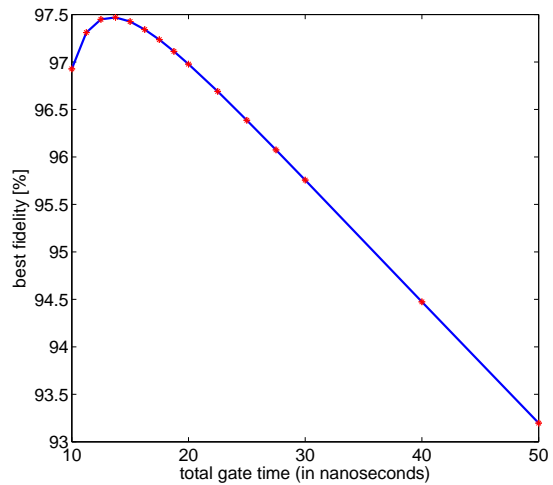}
		\caption{(Color online) Plot of optimal fidelity versus total gate time for Heisenberg interaction in presence of amplitude damping with 500 ns.}
	\label{heisendecoh}
\end{figure}

The origin of the poor fidelity when $\eta$ is small can be understood as follows: In the pulse 
construction (\ref{finalps}) of Ref.~\cite{PhysRevA.81.012320}, a Cartan decomposition 
is used to decompose the time-evolution operator generated by (\ref{hamiltonianrwa}) into
single-qubit rotations, an entangling operator, and a global phase factor. The entangler has 
the form
\begin{equation}
A(x,y,z) \equiv  e^{-i(x \, \sigma_1^x \sigma_2^x +  y \, \sigma_1^y \sigma_2^y
+  z \, \sigma_1^z \sigma_2^z)},
\label{entangler definition}
\end{equation}
where $x,$ $y,$ and $z$ are three coordinates (angles). Following Zhang \textit{et al.}~\cite{ZhangPRA03},
the entangler coordinates trace out a trajectory in the three-dimensional space of entanglers as time
progresses. In the construction of  Ref.~\cite{PhysRevA.81.012320}, the trajectory starts in the plane $x=y,$
and then a refocusing $\pi$ pulse is used to reflect the trajectory to the point $(\frac{\pi}{4},0,0)$ or
$(-\frac{\pi}{4},0,0)$ on $x$ axis. (The actual point reached depends on the sign of $J$.) The time it takes
to do this---neglecting the time needed for the $\pi$ pulse---is $2  \Delta t$ [see (\ref{phidef})], or $\pi \hbar/4g\eta$. Including all the single-qubit rotations in (\ref{finalps}) leads to a total gate time of
\begin{equation} \label{tcnoteta}
t_{\rm gate}=\frac{\pi \hbar}{4g\eta}+\frac{3\pi+2\varphi}{\Omega}.
\end{equation}
Because the first term in (\ref{tcnoteta}) is inversely proportional to $\eta g$, for a fixed gate time
a larger value of coupling strength $g$ is required when $\eta$ is small. But when $g$ increases
the assumption of weak coupling used in \cite{PhysRevA.81.012320} is violated and the corrections 
to the rotating-wave-approximation get larger. Furthermore, that large coupling leads to considerable
errors during the single-qubit operations because the qubit-qubit interaction is not switched off.

Although we have focused on the intrinsic fidelity in this work, it is interesting to calculate one example of a fidelity curve with decoherence. We choose the Heisenberg interaction for this study, with $500 \, {\rm ns}$ amplitude damping. Reoptimizing $\Omega$ and $g$ for each total gate time leads to the fidelity curve shown in Fig.(\ref{heisendecoh}). Table \ref{heisendecohtable} gives the corresponding optimal parameters. The curve exhibits a maximum fidelity ($\approx 97.47\%$) at about 13.75 ns. which represents the optimal time to construct a CNOT with this assumed decoherence model. The optimal interaction strength and Rabi frequency are $g/h \approx 12.89 \, {\rm MHz}$ and $\Omega/h \approx 370 \, {\rm MHz} 
$.

Figs.(\ref{random}) and (\ref{randomavg}) are our principal results. To use the fidelity curve of
Fig.(\ref{randomavg}) for a particular application, one should calculate the $\eta$ value for the
application and extrapolate between the curves provided. We note, however, that for small
$\eta$ the pulse sequence (\ref{finalps}) is not useful, and one should construct an alternative
pulse sequence using the methods of Refs.~\cite{ZhangPRA03} and \cite{PhysRevA.81.012320}
to generate an entangler on the $z$ axis instead of the $x$ axis.

\begin{acknowledgments}
This work was supported by IARPA under grant no. W911NF-04-1-0204. The authors would like to thank Andrei Galiautdinov, John Martinis, and Emily Pritchett for useful discussions.
\end{acknowledgments}

\bibliography{JGqc}

\end{document}